\documentclass[aps, prl, twocolumn, amsmath, amssymb, nofootinbib]{revtex4-2}

\usepackage{amsmath}
\usepackage{mathrsfs}
\usepackage{bm}
\usepackage{braket}
\usepackage{color}
\usepackage{tikz}
\usetikzlibrary{calc}
\usetikzlibrary{positioning}
\usepackage[capitalise, nameinlink]{cleveref}

\makeatletter
\newcommand*{\sumcirclearrowleft}{%
  \DOTSB
  \mathop{
    \mathchoice
      {\rlap{\kern.25em\rotatebox[origin=c]{-90}{$\circlearrowleft$}}{\sum}}
      {\vcenter{\rlap{\kern.2em\rotatebox[origin=c]{-90}{$\scriptscriptstyle\circlearrowleft$}}}{\sum}}
      {\sum}{\sum}
  }\slimits@
}

\newcommand*{\sumcirclearrowright}{%
  \DOTSB
  \mathop{
    \mathchoice
      {\rlap{\kern.25em\rotatebox[origin=c]{90}{$\circlearrowright$}}{\sum}}
      {\vcenter{\rlap{\kern.2em\rotatebox[origin=c]{90}{$\scriptscriptstyle\circlearrowright$}}}{\sum}}
      {\sum}{\sum}
  }\slimits@
}
\makeatother

\begin{document}

\title{Fermionized dual vortex theory for magnetized kagom\'{e} spin liquid}

 \author{Si-Yu Pan$^{1,2}$}
 \author{Gang v.~Chen$^{1,2}$}
 \email{chenxray@pku.edu.cn}
\affiliation{$^{1}$International Center for Quantum Materials, School of Physics, Peking University, Beijing 100871, China}
\affiliation{$^{2}$Collaborative Innovation Center of Quantum Matter, 100871, Beijing, China}
 
\begin{abstract}
Inspired by the recent quantum oscillation measurement on the kagom\'{e} lattice antiferromagnet 
in finite magnetic fields, we raise the question about the physical contents of the emergent fermions 
and the gauge fields if the U(1) spin liquid is relevant for the finite-field kagom\'{e} lattice antiferromagnet. 
Clearly, the magnetic field is non-perturbative in this regime, and the finite-field state has no direct 
relation with the U(1) Dirac spin liquid proposal at zero field. We here consider the fermionized dual 
vortex liquid state as one possible candidate theory to understand the magnetized kagom\'{e} spin liquid. 
Within the dual vortex theory, the $S^z$ magnetization is the emergent U(1) gauge flux, and the fermionized 
dual vortex is the emergent fermion. The magnetic field polarizes the spin component that modulates the 
U(1) gauge flux for the fermionized vortices and generates the quantum oscillation. Within the mean-field theory, 
we discuss the gauge field correlation, the vortex-antivortex continuum and the vortex thermal Hall effect. 
\end{abstract}

\maketitle
 
\noindent{\emph{\bf Introduction.}}---Establishing the connection between the microscopic degrees of freedom 
and the emergent variables in the underlying framework is 
the key step to build up theories and understand the experimental  
phenomena for quantum many-body systems. This is particularly   
so for exotic quantum phases of matter such as quantum spin liquids. 
In quantum spin liquids, for example, one ought to identify the fractionalized 
quasiparticles and the gauge fields in the physical spin variables, 
as the quantum spin liquids are described by the emergent gauge theories 
in their deconfined phases that support the fractionalized spinon-like quasiparticles. 
These connections could provide useful mutual feedback between
theories and experiments.

Quite recently, there is some experimental progress on the kagom\'{e} 
lattice spin liquid candidate material YCu$_3$-Br~\cite{zheng2023unconventional}.  
A 1/9 magnetization plateau was observed in the magnetic field, 
and sets of oscillation emerges in the vicinity of this plateau. 
This was understood from the response of the fermionic 
spinons to the emergent U(1) gauge field. 
In fact, quantum oscillation of spin liquids has been proposed long time ago 
by Lesik Motrunich~\cite{Motrunich2006}.
 The system in Motrunich's analysis is in the weak Mott regime
 and has little resemblance with the strong Mott insulating kagom\'{e} lattice 
 antiferromagnet. Crudely speaking, 
the emergent fermionic spinon in the weak Mott regime is not very far from the physical electron. 
Microscopically, the strong charge fluctuation generates the ring exchange that 
traps the external magnetic flux and then induces the internal U(1) gauge flux
via the scalar spin chirality $({\boldsymbol S}_i \times {\boldsymbol S}_j)\cdot {\boldsymbol S}_k$. 
This induction mechanism is the physical origin of quantum oscillation for the weak 
Mott insulating spin liquid. In the kagom\'{e} lattice spin liquid materials 
that are in the strong Mott regime, Motrunich's mechanism does not apply.   
In the strong Mott regime, if one still adopts the usual fermionic parton/spinon construction and relates
the scalar spin chirality to the U(1) gauge flux for the spinon, the external field does not directly induce
the internal gauge flux via the simple Zeeman coupling. 
 The brute-force attack is to examine 
the spinon-gauge-coupled theory and analyze the energetics for the spin model by varying the magnetic fields,
and then check if the background U(1) gauge flux is modified in such a way to generate the quantum oscillation.

Instead of invoking the involved energetics with the fermionic parton construction 
for the spontaneous internal flux generation where
the actual spin model is {\sl not} quite clear for the material, we return to the origin. 
From the experiments, the quantum oscillation was not observed near 
the zero field where the most attention was drawn and the spin liquid was 
proposed~\cite{RevModPhys.88.041002,PhysRevLett.98.117205,PhysRevB.77.224413}. 
Thus, it is reasonable to expect that, the zero-field spin liquid and the spin liquid 
near the 1/9 magnetization plateau~\cite{Nishimoto_2013} are different states. 
The magnetic fields are non-perturbative near the plateau, 
and the spin symmetry of the Heisenberg model breaks down to U(1). 
It is tempting to think that, a different theoretical framework and description 
from the slave-fermion construction may be needed for the spin liquid 
near the 1/9 plateau. The difference would bring 
a different relation between the physical spin variables
with the fractionalized quasiparticles and gauge fields.

Around the 1/9 plateau, the spin component, $S^z$, is magnetized.   
From the internal perspective of the emergent spinon-gauge theory,  
a finite and varying $S^z$ should be responsible for the generation of the internal U(1) gauge flux. 
Thus, it is tempting for us to regard $S^z$ to be directly related to the internal gauge flux 
for the emergent fermionic matter. 
Due to the U(1) symmetry here, it is convenient to think 
the spin variables in terms of the hardcore bosons 
%with ${S^z_i = n_i -1/2 }$ and $S^+_i = b^\dagger_i, S^-_i = b_i^{}$ \\
where $S^z$ is related to the boson density. 
If $S^z$ is interpreted as the U(1) gauge flux, 
from the traditional formulation of the 
boson-vortex duality, the boson density, i.e. 
$S^z$ directly serves as the dual U(1) gauge flux for the vortices~\cite{Fisher2004}. 
Due to the gauge fluctuations, the vortices are interacting with the logarithmic repulsion. 
With the frustrated spin interaction in the transverse components, the vortices are 
at half-filling~\cite{PhysRevB.72.064407}. Following the existing arguments by M.P.A. Fisher 
and his collaborators~\cite{PhysRevB.72.064407,PhysRevLett.95.247203,PhysRevB.73.174430,PhysRevB.75.184406,PhysRevB.75.144411},
one can infer that, the strong vortex interaction and the finite vortex density 
suppress the density fluctuations of the vortices such that the   
vortex exchange statistics becomes less important. Given the fermionic 
quantum oscillatory phenomena in experiments, one naturally thinks the vortices
as fermions. Formally, one can perform an exact statistical transmission 
by attach the $2\pi$ flux tubes to the bosonic vortex~\cite{PhysRevB.72.064407}. 
This exact formulated theory is the starting point for the mean-field theory 
that describes the low-energy fermionic vortices coupled to the dual U(1) gauge field. 
In this paper, we build up a fermionized dual vortex theory as a possible understanding for the 
magnetized kagom\'{e} spin liquid. In this theoretical description, the emergent fermions 
are the fermionized vortices, and the emergent gauge field is the dual U(1) gauge field. 
The purpose of this work is to make the specific connection between the recent puzzling 
quantum oscillation experiments and the fermionized dual vortex theory, 
and further identify the physical properties within the dual vortex theory.

\noindent{\emph{\bf Duality formulation for the spin model.}}---To formulate the theoretical description, 
we start from the antiferromagnetic spin model on the kagom\'{e} lattice in the external magnetic field, 
\begin{eqnarray}
    H &=& \sum_{ij} \big[ J_{z,ij} S^z_i S^z_j + \frac{1}{2} {J_{\perp,ij} } (S^+_i S^-_j + S^-_i S^+_j) \big] 
    \nonumber \\
                          &&      - \sum_i  h \, S^z_i. 
    \label{eq1}
\end{eqnarray}
To keep the model a bit more general, we set the the spin model as the generic spin-1/2 XXZ model in the magnetic field
for which the nearest-neighbor Heisenberg model is included as the limited case. 
 As the theory in this work is not a study of the energetics, 
we simply make sure the model retains the required symmetries, 
especially the U(1) symmetry, for the construction purpose. 
In the hardcore boson representation for the spin variable, one sets
\begin{eqnarray}
S^z_i = n_i - \frac{1}{2},   \quad S^+_i = b^\dagger_i, \quad S^-_i = b_i^{} ,
\end{eqnarray}
where $n_i$ is the boson density and $b^\dagger_i$ ($b_i$)
is the creation (annihilation) operator for the boson. 
Due to the hardcore constraint, $n_i=0,1$ for the spin-1/2 variables.  
The spin model now becomes
\begin{eqnarray}
H &=& \sum_{ij} \big[  J_{z,ij} (n_i - \frac{1}{2} ) (n_j - \frac{1}{2} )  + \frac{J_{\perp,ij} }{2} (b^\dagger_i b_j +h.c.) \big]
\nonumber \\
&& -\sum h (n_i - \frac{1}{2}). 
\end{eqnarray}

It is more convenient to replace the hardcore bosons with 
the bosonic rotors by relaxing the hardcore constraint and 
introducing the onsite Hubbard-$U$ interaction to suppress 
the boson density. In the bosonic rotor representation, one 
has $b_i \sim e^{-i \phi_i}$, and the above model is converted to 
\begin{eqnarray}
H &=& \sum_{ij}  J_{\perp,ij} \cos (\phi_i - \phi_j) + \sum_i U (n_i - \bar{n})^2 
\nonumber \\
&& +   \sum_{ij} J_{z,ij} (n_i - \bar{n})(n_j - \bar{n}) + \cdots,
\label{eq2}
\end{eqnarray}
where the phase variable $\phi_i$ is conjugate to the boson density    
$n_i$ with ${[\phi_i, n_j] = + i\delta_{ij}}$. 
The onsite interaction $U$ is a strong interaction 
that is introduced to implement the Hilbert space constraint for the hardcore bosons
to enforce the selection of $n_i=0, 1$. 
For the zero-magnetization 
state with $\langle S^z \rangle=0$, the boson is at half-filling $\bar{n}=1/2$. 
For the 1/9-magnetization plateau state that is of interest in this work, 
we have ${\bar{n} =1/2+\langle S^z\rangle =  5/9}$.

\begin{figure}[b]
\includegraphics[width=5.5cm]{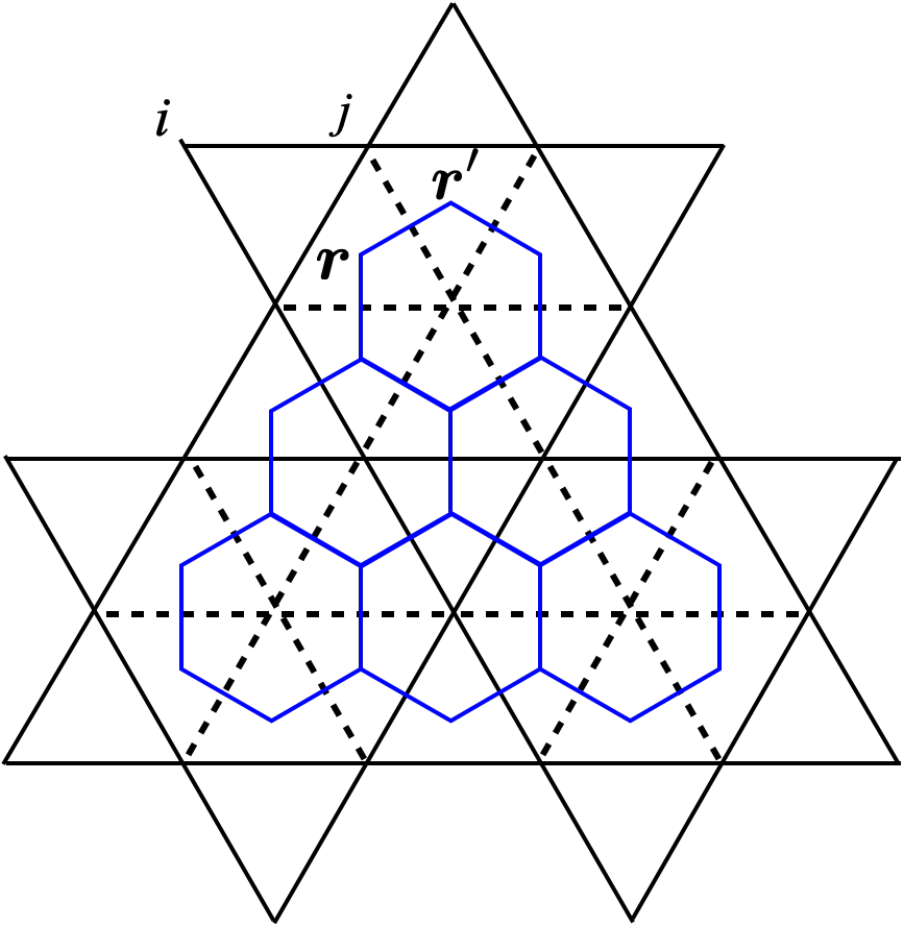}
\caption{(Color online.) The interpolated lattice between the kagom\'{e} lattice and the triangular lattice. 
When the interactions on the dashed bonds are tuned to zero, the system becomes a kagom\'{e} lattice.
Otherwise, it is a triangular lattice. 
The introduced sites are referred as the auxiliary sites. 
To distinguish the interactions on the solid bonds, the exchange couplings on the dashed bonds
are referred as $J'_{\perp}$ and $J'_z$. 
The dual lattice of the triangular lattice is a honeycomb lattice (in blue). } 
\label{fig1}
 \end{figure}

The procedure of the boson-vortex duality is quite standard. 
To manifest the vortex degrees of freedom explicitly, 
one performs the duality transformation. The resulting model describes the 
mobile vortices hop on the lattice sites (${\boldsymbol r}, {\boldsymbol r}'$) 
of the dual lattice in the background of the fluctuating U(1) gauge field 
$a_{{\boldsymbol r}{\boldsymbol r}'}$. The background U(1) gauge flux, 
that is experienced by the vortices, arises from the boson density and/or
the magnetization, i.e.
\begin{eqnarray}
 S^z_i \sim n_i \sim \frac{1}{2\pi} (\Delta \times a)_{i} . 
\label{eq3}
\end{eqnarray}
The above linear relation between the magnetization and the internal gauge flux
encodes the underlying reason for the quantum oscillation within this framework
that has been previously argued in the introduction and will be further discussed below. 
A similar linear gauge-flux induction and the resulting flux-matter 
coupling~\cite{PhysRevResearch.2.013066,PhysRevB.96.195127} 
have been mentioned in the context of the pyrochlore quantum ice U(1) spin liquid
except that the matter fields over there are bosonic~\cite{PhysRevB.69.064404}. 
No quantum oscillation, however, is expected for the pyrochlore U(1) spin liquid with 
the bosonic matters.

\begin{figure}[t]
\includegraphics[width=8.6cm]{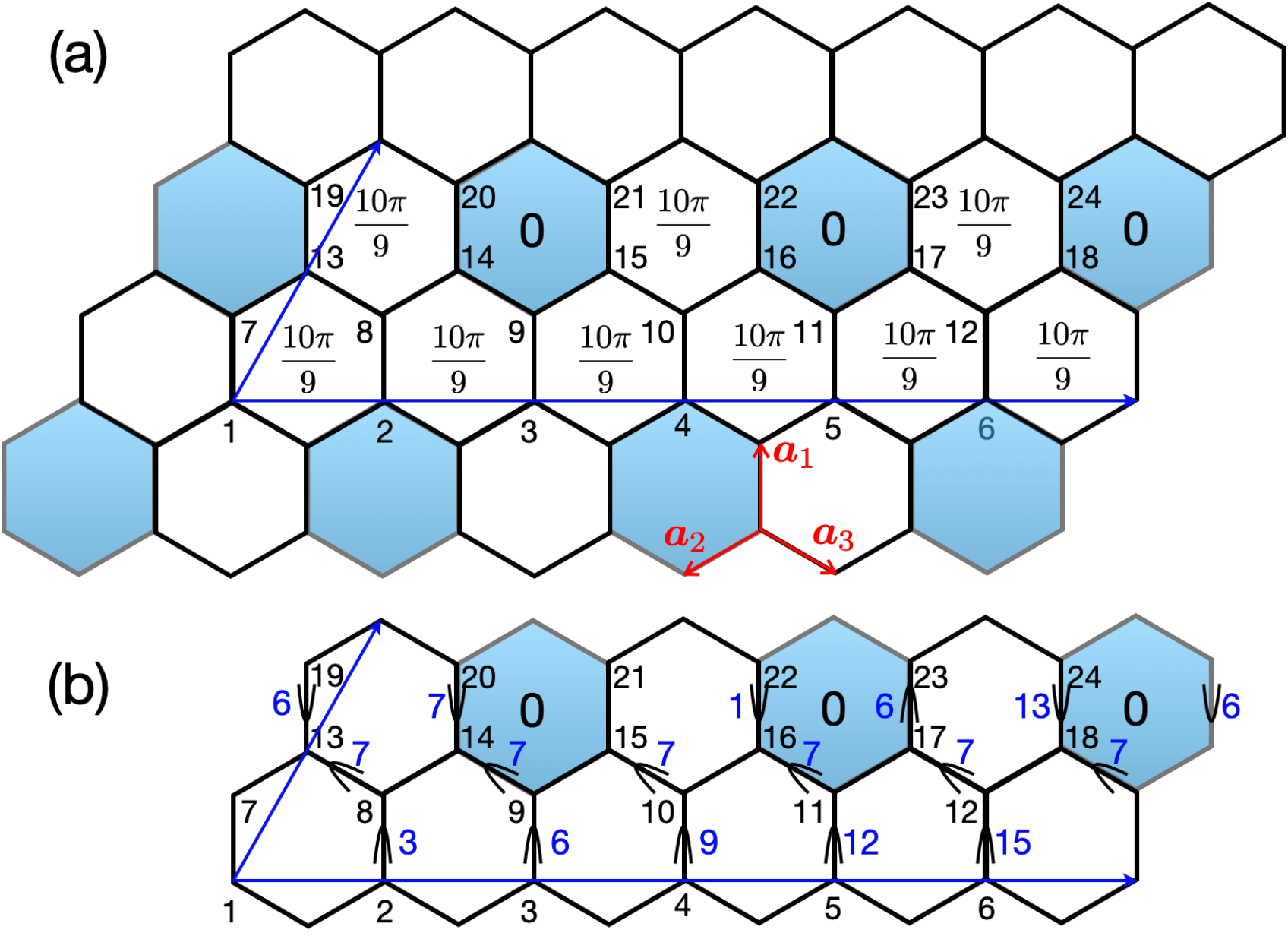}
\caption{(Color online.)
(a) The U(1) gauge flux distribution on the dual honeycomb lattice. 
The blue arrows define the enlarged unit cell once the gauge fixing procedure is implemented. 
The blank (blue) hexagon refers to $10\pi/9$ (0) background U(1) gauge flux. 
The numbers 1,2, $\cdots$ 24 are the sublattice indices after the gauge fixing.
The nearest-neighbor vectors ${\boldsymbol a}_1, {\boldsymbol a}_2, {\boldsymbol a}_3$ are set to unity.   
(b) 
The background U(1) phase $\bar{a}_{ij}$ of each bond is given as $2\pi {m_{ij}/18}$,
where $m_{ij}$ is the blue integer on the bond $ij$ in the plot, and the arrow is the direction of gauge link. 
The unmarked bonds have $m_{ij} =0$.
}
\label{fig2}
\end{figure}

The dual lattice of the kagom\'{e} lattice is a dice lattice and contains three sublattices
with different coordination numbers. From the previous experiences~\cite{PhysRevB.75.184406}, 
the dual vortex theory on the dice lattice is a bit difficult to deal with when the fermionization 
procedure is introduced. Instead, a useful trick that circumvents the difficulty is to introduce 
the integer spin moments with $\langle S^z \rangle =0$ in the centers of the hexagonal 
plaquettes~\cite{PhysRevB.75.184406}. As shown in Fig.~\ref{fig2}, 
these auxiliary integer moments interact with the nearby spins 
with the weak antiferromagnetic interactions, $J'_{\perp}$ and $J'_z$, on the dashed bonds.  
The hardcore boson mapping on these auxiliary sites yields $n_i = S^z_i$ and 
$\bar{n}_i = \langle S^z_i \rangle =0$. This boson-vortex duality then 
yields the dual vortex model on the dual honeycomb lattice. 
The dual Hamiltonian on the dual honeycomb lattice now has the following form, 
 \begin{eqnarray}
H_{\text{dual}} &=&  -\sum_{\langle {\boldsymbol r} {\boldsymbol r}' \rangle} 
t_{ {\boldsymbol r} {\boldsymbol r}'}  \cos ( \theta_{\boldsymbol r} - \theta_{{\boldsymbol r}'} 
- \bar{a}_{{\boldsymbol r}{\boldsymbol r}'} -  a_{{\boldsymbol r}{\boldsymbol r}'}) 
\nonumber \\
&& 
+ \sum_{ {\boldsymbol r} {\boldsymbol r}' } (N_{\boldsymbol r} -\frac{1}{2} ) 
V_{{\boldsymbol r} {\boldsymbol r}'} (N_{\boldsymbol r'} -\frac{1}{2} )  
\nonumber \\
&& 
+  \sum_{{\boldsymbol r} {\boldsymbol r}'   }   2\pi^2
    J_{{\boldsymbol r} {\boldsymbol r}'}^{} e^2_{{\boldsymbol r} {\boldsymbol r}'}
+\sum_{\boldsymbol r}   \frac{U}{(2\pi)^2}  (\Delta \times a)_{\boldsymbol r}^2 .
\label{eq4} 
\end{eqnarray}
Here ${\boldsymbol r}, {\boldsymbol r}'$ refer to the sites of the dual honeycomb lattice in Fig.~\ref{fig1}.
The phase variable $\theta_{\boldsymbol r}$ is conjugate with the vortex density $N_{\boldsymbol r}$
with
\begin{equation}
[ \theta_{\boldsymbol r}, N_{{\boldsymbol r}'}] =  i\delta_{ {\boldsymbol r}{\boldsymbol r}'}, 
\end{equation}
such that $e^{\pm i \theta_{\boldsymbol r}}$ creates/annihilates the vortex at the lattice 
${\boldsymbol r}$. When the vortices hop on the dual lattice, they are coupled to 
the underlying U(1) gauge field $a_{{\boldsymbol r}{\boldsymbol r}'}$ 
that is defined on the link of the dual lattice. 
In the first line of Eq.~\eqref{eq4}, $a_{{\boldsymbol r}{\boldsymbol r}'}$ is the fluctuating piece of the gauge 
field, while $\bar{a}_{{\boldsymbol r}{\boldsymbol r}'}$ is the background gauge field to take care of the finite 
boson density. 
The second line Eq.~\eqref{eq4} describes the interaction between the vortices from the gauge fluctuations.
The third line of Eq.~\eqref{eq4} describes the standard Maxwell's terms for the U(1) lattice gauge theory,
where the variable ${e}_{{\boldsymbol r}{\boldsymbol r}'}$ defines the ``electric field'' for the gauge field on the link. 
The coupling $J_{{\boldsymbol r}{\boldsymbol r}'}$ is given by $J_{{\boldsymbol r}{\boldsymbol r}'} = J_{\perp} $($J'_{\perp}$)
such that the link ${\boldsymbol r}{\boldsymbol r}'$ on the dual honeycomb lattice  
crosses the original triangular lattice link $\langle ij \rangle$ with $J_{\perp,ij} = J_{\perp}$ ($J'_{\perp}$).

Due to the inhomogeneous boson distribution on the triangular lattice in Fig.~\ref{fig1}, the background U(1) gauge flux 
also receives an inhomogeneous distribution with
\begin{eqnarray}
(\Delta \times \bar{a})_{i} &\equiv & \sumcirclearrowleft_{ \langle {\boldsymbol r}{\boldsymbol r}' \rangle_i} a_{{\boldsymbol r}{\boldsymbol r}'}  
=  2\pi \bar{n}_i 
\nonumber \\
&=& \left \{ 
\begin{array}{cl}
\frac{10\pi}{9}, & i \in \text{kagom\'{e} lattice}
\vspace{1mm}\\
0, & i \in \text{auxiliary sites}
\end{array}
\right.,
\end{eqnarray}
where $\langle {\boldsymbol r}{\boldsymbol r}' \rangle_i$ refers to the dual link around 
the original lattice site $i$.
The U(1) gauge flux distribution for the vortices is depicted in Fig.~\ref{fig2}(a),
and a gauge choice is given in Fig.~\ref{fig2}(b).  
Moreover, the inhomogeneous spin exchange also transfers to the vortex hopping
on the dual lattice such that one would crudely have ${t' > t}$ because ${J'<J}$ allows
an easier vortex tunneling where $t'$ is the vortex hopping on the bonds of the blue 
hexagon and $t$ is the hopping on the remaining bonds. So far, the vortices are bosonic, 
and are strongly interacting. Due to the geometric frustration of the spin model, 
the vortices are at finite density with a half filling per site.

%%%%%%%%%

\noindent{\emph{\bf Fermionized vortex theory.}}---The boson-vortex duality resolves the internal U(1) gauge flux generation via the magnetization,
but has not resolved the emergence of the fermions for the quantum oscillation. This section aims 
to perform the fermionized vortex treatment based on the existing argument 
due to the strong vortex interaction. This procedure amounts to attaching $2\pi$
U(1) gauge flux to each vortex via the Chern-Simons gauge field $A$
with the condition
\begin{eqnarray}
(\Delta \times A)_{\boldsymbol r} = 2\pi N_{\boldsymbol r}. 
\end{eqnarray}
With the attached flux, the vortex becomes fermionic. 
The resulting kinetic part of the fermionized vortices is then given as
\begin{eqnarray}
H_{\text{ferm}} = -\sum_{ \langle {\boldsymbol r}{\boldsymbol r}' \rangle} 
                                          t_{{\boldsymbol r}{\boldsymbol r}'}  
                                             d^{\dagger}_{{\boldsymbol r}'} d^{}_{\boldsymbol r} 
                                                 e^{-i \bar{a}_{{\boldsymbol r}{\boldsymbol r'}} 
                                                 -i a_{{\boldsymbol r}{\boldsymbol r'}} 
                                                 -i A_{{\boldsymbol r}{\boldsymbol r'} } },
\end{eqnarray}
where the fermion operator $d^{\dagger}_{\boldsymbol r}$ ($d^{}_{\boldsymbol r}$) 
creates (annihilates) a fermionized vortex at the site ${\boldsymbol r}$. 
The fermionized vortex is minimally coupled to the dual U(1) gauge field 
via the duality transformation and the Chern-Simons gauge field $A$ via the flux attachment. 
The formulation is exact at this stage.

%%%%%%%%%%%%%%%%

\begin{figure}[t]
\includegraphics[width=8.6cm]{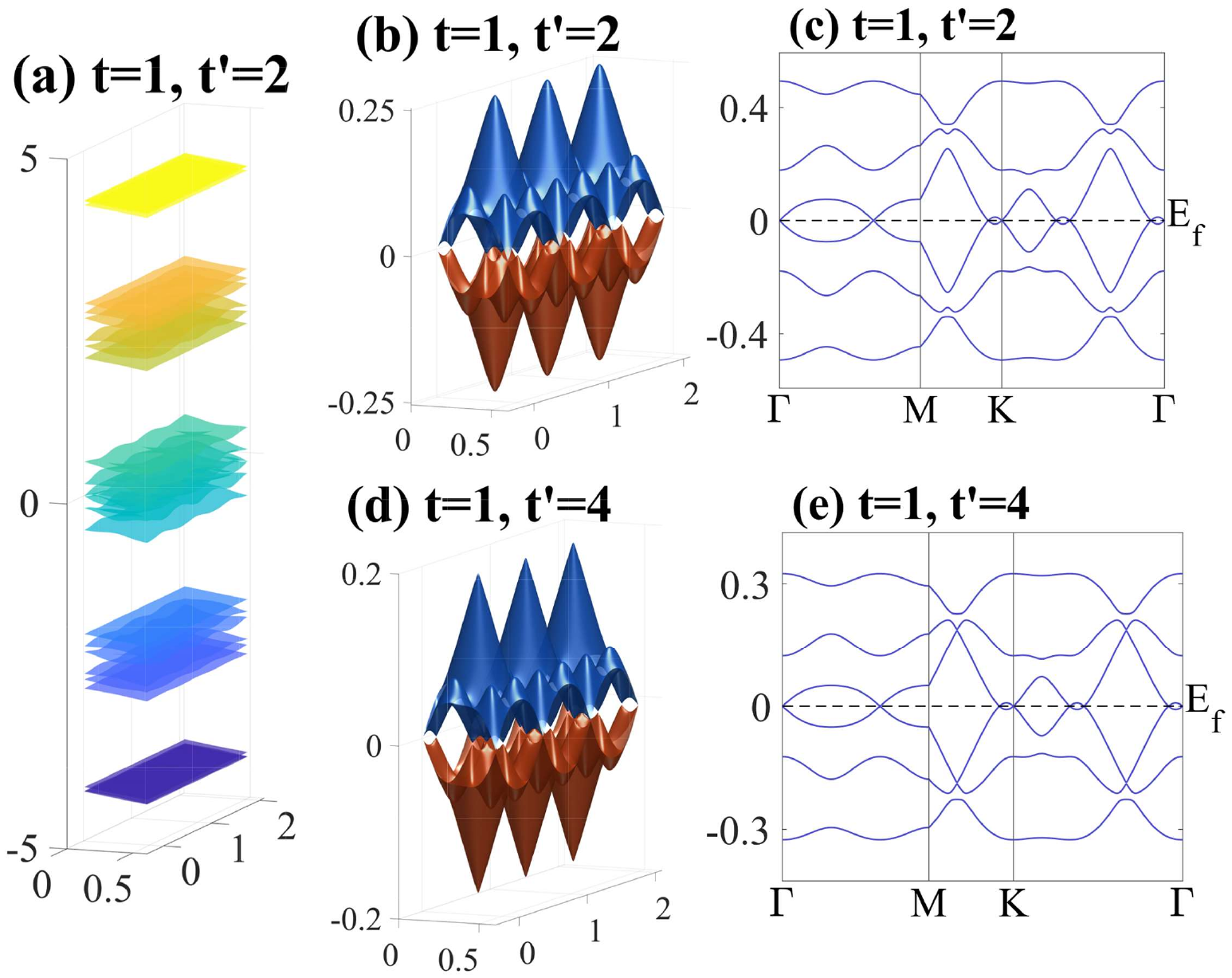}
\caption{(Color online.) The band structure of the fermionized vortices.
 (a) All the bands for the fermionized vortices where we choose with $ {t'=2t}$. 
 (b)(d)The Dirac band touchings at the Fermi energy for $t'=2t$ and $4t$.
 (c)(e) The bands near the Fermi energy for $t'=2t$ and $4t$ along the $\Gamma$-M-K-$\Gamma$ momentum line in the 
 Brillouin zone.}
    \label{fig3}
\end{figure}

\begin{figure}[t]
\includegraphics[width=8.6cm]{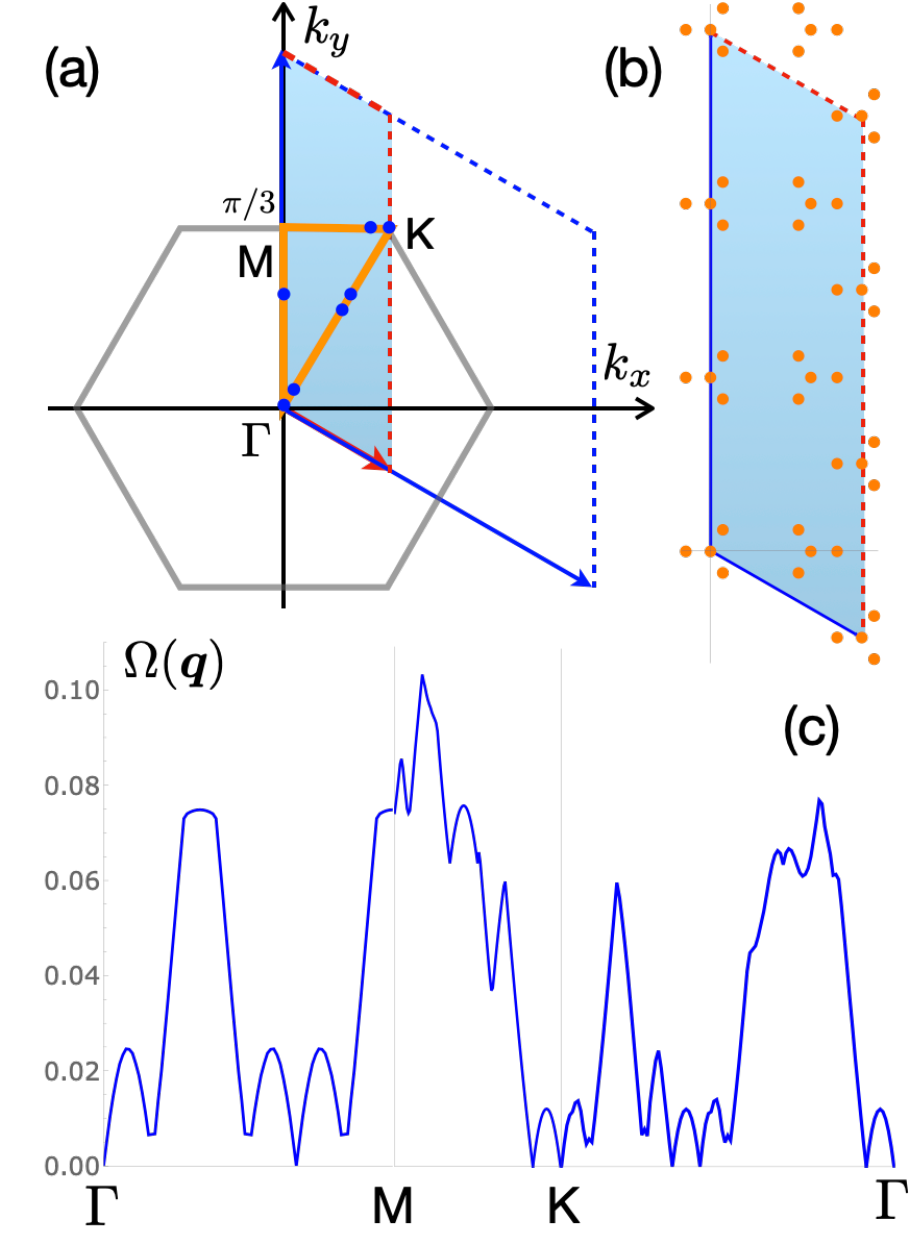}
    \caption{(a) The Brillouin zone of the kagom\'{e} lattice and the reduced
    Brillouin zone (in blue) of the dual honeycomb lattice with the gauge choice in Fig.~\ref{fig2}(b). 
    (b) The enlarged version of the reduced Brillouine zone of (a). The oranges dots are the vortex Dirac points at
   the Fermi level. 
    (c) The lower excitation edges of the vortex-antivortex continuum 
    along $\Gamma$-M-K-$\Gamma$ for ${t'=2t}$, computed
    for $120\times 180$ momentum points in the reduced Brillouin zone.
    The energy unit is $t$. The blue points along $\Gamma$-M-K-$\Gamma$ in (a) are the gapless points in (c). 
    }
    \label{fig4}
\end{figure}

To analyze the interacting fermionized vortex theory, we consider a mean-field
or saddle-point configuration for the U(1) gauge fields and the Chern-Simons gauge fields.
It is straightforward to incorporate the
background U(1) gauge flux via $\bar{a}_{{\boldsymbol r}{\boldsymbol r}'}$ in Fig.~\ref{fig2}(b).
The attached $2\pi$ flux to each vortices is incorporated through a flux-smearing procedure~\cite{PhysRevB.73.174430}.
One first smears the attached $2\pi$ gauge flux of the vortex on the dual lattice to the neighboring 
three hexagons. The total smeared flux from all six sites on each honeycomb 
turns out to be 0 (mod $2\pi$). 
One can then set ${A_{{\boldsymbol r}{\boldsymbol r}'}\rightarrow 0}$ 
in the flux-smeared mean-field treatment. The mean-field theory for the fermionized vortices 
is now written as, 
\begin{eqnarray}
H_{\text{ferm,MF}} =  -\sum_{ \langle {\boldsymbol r}{\boldsymbol r}' \rangle} 
                                          t_{{\boldsymbol r}{\boldsymbol r}'}  
                                             d^{\dagger}_{{\boldsymbol r}'} d^{}_{\boldsymbol r} 
                                               e^{-i \bar{a}_{{\boldsymbol r}{\boldsymbol r'}}} . 
\end{eqnarray}
With the enlarged magnetic unit cell due to the background gauge flux 
in Fig.~\ref{fig2}, there exist 24 bands, $\omega_{\mu} ({\boldsymbol k})$ 
(${\mu=1,2,\cdots, 24}$), for the dual vortices in the reduced Brillouin zone. 
At the half filling, the Fermi surface occurs at a few discrete Dirac points  
that are depicted in Fig.~\ref{fig3} and Fig.~\ref{fig4}. 
The vortex bands near the Fermi level are further plotted along the $\Gamma$-$M$-$K$-$\Gamma$ momentum
 lines in the reciprocal space in Fig.~\ref{fig3}. Due to the deconfinement and the fractionalization,
 the single vortex excitation is not detectable in the spectroscopic measurements. 
 Instead, the gapless Dirac vortex excitations give rise to a $T^2$ specific heat
 and contribute to the thermal conductivity $\kappa_{xx}$.  
 Via the disorder scattering~\cite{PhysRevB.62.1270}, the Dirac vortices are expected to 
 give a finite $\kappa_{xx}/T$ at low temperatures.   
 
\noindent{\emph{\bf Dynamic spin structure factors.}}---As we have previously mentioned,
 the vortex band cannot be directly measured alone. 
In a spectroscopic measurement such as NMR and inelastic neutron scattering 
that will be discussed below, the vortices appear as the vortex-antivortex pair 
excitations in the spectrum. Due to the emergent gapless Dirac fermions near the Fermi level,  
it is illuminating for us to compute the the vortex-antivortex continuum as the physical observable first. 
Later on, we will connect with the spin correlation and the inelastic neutron scattering measurement.

We first reveal the lower excitation edges $\Omega ({\boldsymbol q})$ of the vortex-antivortex
continuum as
\begin{equation}
\Omega ({\boldsymbol q}) = \min_{\boldsymbol k} 
\big[ \omega_{\mu} ({\boldsymbol k} + {\boldsymbol q}) - \omega_{\nu}  ({\boldsymbol k} ) \big],
\end{equation}
where $\omega_{\mu} ({\boldsymbol k})$ ($\omega_{\nu} ({\boldsymbol k})$) is the unfilled (filled) 
vortex dispersion. 
In Fig.~\ref{fig4}, we plot the lower excitation edge of the vortex-antivortex continuum for $t'=2t$.
We have computed other parameters, and the structures are not quite different except the overall 
energy scales. The inter- and intra-Dirac-cone scattering events are most visible as the gapless 
excitations at various momentum points that are marked along the path $\Gamma$-M-K-$\Gamma$ in Fig.~\ref{fig4}(a). 
These gapless modes are important information for the experimental or numerical identification.

We return to the spin correlation functions that are directly measured in the inelastic neutron scattering experiments. 
The $S^z$ operator, not only relates to the internal gauge flux, but also has the contribution from the vortex loop currents. 
The former indicates that, the $S^z$-$S^z$ correlation 
contains an important piece from the U(1) gauge field correlation. 
Despite this useful relation, the gauge field ($\Delta \times a$) correlation suffers from the usual suppression of intensity
at low energies~\cite{LeeNagaosa2013,PhysRevB.75.184406}. As a comparison, for the scalar spin chirality 
description of the gauge flux in the slave-fermion construction, the $S^z$-$S^z$ correlation cannot be 
directly related to the gauge field fluctuations. With the assistance from the Dzyaloshinskii-Moriya interaction,
there can be a piece of gauge field fluctuation in the $S^z$-$S^z$ correlation, but is further suppressed 
by $\sim \mathcal{O}(D_z/J)^2$ where $D_z$ is the strength of the out-of-plane Dzyaloshinskii-Moriya interaction. 
The latter contribution from the vortex loop current to $S^z$ indicates the presence of the vortex-antivortex
continuum in the $S^z$-$S^z$ correlation. Thus, one could examine the positions of the gapless excitations 
and other structures from the lower excitation edge of the $S^z$-$S^z$ dynamic spin structure factor.

The $S^+$ operator changes the $S^z$ quantum number and thus creates $2\pi$ gauge flux. 
Classically, with the introduced $2\pi$ background flux, each Dirac fermion 
has the quasi-localized zero-energy state near the introduced flux at the zero energy limit.
A detailed symmetry analysis is required to establish the relation between 
 $S^+$ and the monopole insertion operators in the low-energy field theory 
 together with their momenta~\cite{PhysRevB.73.174430}, and this may be studied in the future.

\noindent{\emph{\bf Discussion.}}---The Dirac spectrum of the fermionized vortices immediately leads to the prediction of 
specific heat ${C_v \sim T^2}$, and the thermal conductivity ${\kappa \sim T}$ at low temperatures~\cite{PhysRevB.46.5621}. 
As the variation of the internal U(1) flux also generates the Berry curvature distribution of 
the relevant quasiparticles, quantum oscillation often implies the existence of thermal Hall effect,
though the reverse is not true~\cite{zhang2023thermal}. Here, due to the fractional flux (especially when the flux is varied), 
the vortex bands would generically have a finite Berry curvature distribution and give rise to 
the thermal Hall effect. But this is not supposed to be a surprising nor unique effect. 
What can be a useful and important indication from the thermal Hall measurement
is the emergent fermionic excitations. It was suggested that~\cite{PhysRevLett.124.186602}, 
$\kappa_{xy}/T$ could show universal behaviors (in the temperature dependence) at the finite temperatures that 
are related to the statistics of the emergent excitation. The fermionized vortices 
are the energy carriers, and if they persists to finite temperatures, such a universal behavior
could be useful. It may not confirm the fermionized vortices, but confirm the emergent fermionic 
statistics in the strongly interacting spin systems.

With the background U(1) gauge flux,
the magnetic unit cell is three times of the original unit cell, and the translation symmetry 
is realized projectively on the fermionized vortices for the flux-smeared mean-field state. 
The vortex-antivortex continuum would exhibit the translation symmetry fractionalization 
with the spectral periodicity enhancement~\cite{PhysRevB.90.121102}. 
Unfortunately, the crystal unit cell of the underlying material
YCu$_3$-Br~\cite{zheng2023unconventional} already exhibit tripling compared to the kagom\'{e} lattice. 
So this is probably not a useful signature for this system, 
but applies to other systems and/or the numerical calculation of the spin dynamics. 
Nevertheless, the rich structure of the spin excitation from the fermionized vortex theory 
is quite useful for the further study.

About the candidate material YCu$_3$-Br, the candidate zero-field spin liquid is expected 
to be quite different from the spin liquid proposal for the kagom\'{e} lattice
Heisenberg model and the herbertsmithite~\cite{PhysRevLett.98.117205,RevModPhys.88.041002}. 
The positions of scattering intensity, that are equivalent to the K points~\cite{zeng2023dirac}, differ from the
M points expected from Dirac spin liquid of Ref.~\onlinecite{PhysRevLett.98.117205}, 
algebraic vortex liquid of Ref.~\onlinecite{PhysRevB.75.184406},
and the minima of gapped spin excitations in $\mathbb{Z}_2$ 
spin liquids~\cite{PhysRevB.45.12377,PhysRevB.74.174423}. 
This distinction might lie in the different structures and/or different 
spin models of the material. The puzzle is that, the Heisenberg spin model in the field gives 
rise to 1/9 magnetization plateau~\cite{Nishimoto_2013}, but the Heisenberg spin model at zero field 
does not seem to give rise to the K point scattering intensity. 
Although our proposal in this work does 
not strongly depend on the model, the actual spin model for YCu$_3$-Br 
needs a revisit.

In summary, the current work is inspired from the recent quantum oscillation result in the magnetized
kagom\'{e} spin liquid and attempts to propose one interpretation from the fermionized dual vortex theory. 
In this interpretation, the magnetization serves as the emergent gauge flux, and the emergent fermions 
are the fermionized vortices. This interpretation provides a different induction mechanism of the internal 
U(1) orbital flux through the external magnetic fields from the Motrunich's weak Mott insulating case. 
In the future, more theoretical analysis are needed for the low-energy effective theory of the U(1) Dirac vortex 
liquid state, and the more quantitative connection to the experiments. The proposed state here in this work 
merely takes one step along the line of dual vortex theory and 
is the starting point for further analysis and mutual feedback with experiments.

\noindent{\emph{\bf Acknowledgments.}}---We acknowledge Matthew Fisher and Jason Alicea for communication, 
and Jiawei Mei for discussion. This work is supported by the Ministry of Science and Technology of China with Grants
No.~2021YFA1400300, and the National Science Foundation of China with Grant No.~92065203, 
and by the Fundamental Research Funds for the Central Universities, Peking University. 

\bibliography{refs.bib}

\end{document}